\documentclass[amsmath,amssymb,11pt,superscriptaddress,reprint, preprintnumbers, notitlepage,aps,prl,twocolumn, dvipsnames,table]{revtex4-2}
\pdfoutput=1 
\usepackage[utf8]{inputenc}
\usepackage[english]{babel}
\usepackage{amsmath}
\usepackage{graphicx}
\usepackage{dcolumn}
\usepackage{pbox}
\usepackage{amssymb}
\usepackage{epsfig}
\usepackage[dvipsnames]{xcolor}
\usepackage[normalem]{ulem}
\usepackage{slashed}
\usepackage{amssymb}
\usepackage{verbatim} 
\usepackage{ mathrsfs }
\usepackage{color}
\usepackage[font=small]{caption}
\usepackage[font=small]{subcaption}
\usepackage{url}
\usepackage{tikz}
\usepackage[compat=1.1.0]{tikz-feynman}
\definecolor{MyDarkBlue}{rgb}{0.1, 0.1, 0.8}
\definecolor{SBlue}{rgb}{0.2, 0.4, 0.7} 
\definecolor{MyLightBlue}{rgb}{0.22,0.51,0.9}
\definecolor{MyGreen}{rgb}{0.0, 0.5, 0.0}
\definecolor{BrickRed}{rgb}{0.8, 0.25, 0.33}
\RequirePackage{hyperref}
\hypersetup{colorlinks, citecolor=SBlue,linkcolor=MyDarkBlue, urlcolor=PineGreen}
\usepackage[toc,page]{appendix}

\usepackage{amsmath}
\usepackage{amssymb}
\usepackage{array}
\usepackage{multirow}
\usepackage{booktabs}
\usepackage{xcolor}

\makeatletter

\makeatletter
\renewcommand\@makecaption[2]{%
  \par
  \vskip\abovecaptionskip
  \begingroup
  
   \small\rmfamily
    \begingroup
     \samepage
     
     \flushing
     \let\footnote\@footnotemark@gobble
     \@make@capt@title{#1}{#2}\par
    \endgroup
  \endgroup
  \vskip\belowcaptionskip
}
\makeatother

\DeclareUnicodeCharacter{2212}{-}
\usepackage{orcidlink}

\makeatletter
\@booleanfalse\footinbib@sw
\makeatother

\begin{document}

\preprint{HRI-RECAPP-2026-06}

\title{Real versus Complex Fermion Bare Mass and Multi-Wall String Defect Annihilation}

\author{Utsav Atta}
\email{utsavatta@hri.res.in}
\affiliation{Harish-Chandra Research Institute, Chhatnag Road, Jhunsi, Prayagraj 211019, India}
\affiliation{Homi Bhabha National Institute, Training School Complex, Anushakti Nagar, Mumbai 400094, India}

\author{Tathagata Ghosh\,\orcidlink{0000-0002-8259-0328}}
\email{tathagataghosh@hri.res.in}
\affiliation{Harish-Chandra Research Institute, Chhatnag Road, Jhunsi, Prayagraj 211019, India}
\affiliation{Homi Bhabha National Institute, Training School Complex, Anushakti Nagar, Mumbai 400094, India}

\author{Sudip Manna\,\orcidlink{0009-0000-2162-0506}}
\email{sudipmanna@hri.res.in}
\affiliation{Harish-Chandra Research Institute, Chhatnag Road, Jhunsi, Prayagraj 211019, India}
\affiliation{Homi Bhabha National Institute, Training School Complex, Anushakti Nagar, Mumbai 400094, India}

\begin{abstract}

Particle physics models beyond the Standard Model often contain global symmetries to address various unanswered questions. However, a common criticism of theories based on global symmetries is that such symmetries are generally expected to be explicitly violated by gravitational effects at the Planck scale. In the case of a global $U(1)$ symmetry, this explicit breaking can reduce the symmetry to a discrete subgroup of $U(1)$, leading to the formation of cosmic strings attached to multiple domain walls (DWs). These DWs are usually cosmologically problematic, since their slow scaling behavior can eventually dominate the energy density of the Universe, giving rise to the well-known cosmological DW problem, which is strongly constrained by Big Bang Nucleosynthesis. In this work, we propose a mechanism for the annihilation of these DWs. If a fermion coupled to the symmetry-breaking scalar has a small bare mass, radiative corrections generate a temperature-dependent vacuum bias that drives DW collapse. As an example, we consider a majoron model with right-handed neutrinos, where gravity-induced operators produce a string-wall network. We show that a real bare Majorana mass is sufficient to induce complete DW annihilation for even-dimensional gravity-induced operators, whereas odd-dimensional operators require a complex bare Majorana mass whose phase lifts the residual vacuum degeneracy, enabling complete annihilation.
\end{abstract}

\maketitle
\textbf{\emph{Introduction}.--} Extensions of the Standard Model (SM) often introduce additional scalar field(s) and global symmetries under which the scalar fields carry nontrivial charges. Once these scalar(s) acquire nonzero vacuum expectation values (vevs) and spontaneously break the corresponding global symmetries, several interesting cosmological consequences can arise. In particular, when such spontaneous symmetry breaking (SSB) occurs in the early Universe, the vacuum manifold may possess nontrivial topology, leading to the formation of different topological defects such as cosmic strings~\cite{Kibble:1976sj,Vilenkin:1982ks}, domain walls (DWs)~\cite{Bogomolny:1975de,Vilenkin:1982ks}, monopoles~\cite{tHooft:1974kcl,Polyakov:1974ek}, and hybrid configurations including monopoles connected by strings~\cite{Nambu:1977ag,Kibble:2015twa}, string attached with DWs~\cite{Vilenkin:1982ks,Rothstein:1992rh}, and others~\cite{Turok:1989ai,Bennett:1990xy}.

These topological defects can have important cosmological implications and are subject to various observational and theoretical constraints. Among them, DWs are particularly constrained since their energy density redshifts more slowly than that of radiation and matter. As a result, a stable DW network can eventually dominate the energy density of the Universe, drastically altering the standard cosmological evolution, thereby giving rise to the well known cosmological DW problem~\cite{Zeldovich:1974uw}. For instance, long-lived DWs would be in tension with the primordial light-element abundances inferred from BBN observations. Consequently, viable particle physics models that predict DW formation must also provide mechanisms that ensure their annihilation before the temperature of the Universe drops below $T_{\rm BBN}\sim \mathcal{O}(1\,\mathrm{MeV})$.

Another important ingredient relevant to this discussion arises from quantum gravitational effects. It has long been argued that such effects explicitly break exact global symmetries at the Planck scale. In particular, Ref.~\cite{Abbott:1989jw} argued that exotic spacetime structures such as wormholes are expected to violate global symmetries through the generation of Planck-suppressed operators. Such effects can therefore significantly influence the cosmological evolution of topological defects formed after the spontaneous breaking of these global symmetries in the early Universe.

\begin{figure}
    \centering
    \includegraphics[width=0.48\textwidth]{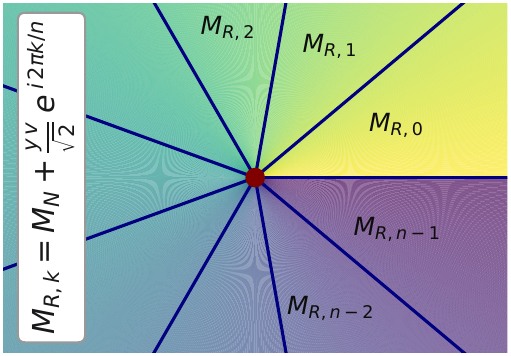}
   \caption{Schematic cross-sectional view of a cosmic string (red blob) attached to $n$ domain walls (blue solid lines), illustrating how the fermion (right-handed neutrino) mass varies across different quadrants due to it's non-zero bare mass term.}
    \label{fig:BLscale}
\end{figure}

In this work, we study the annihilation dynamics of DWs in theories with a continuous global $U(1)$ symmetry. In such scenarios, gravity-induced Planck-suppressed operators explicitly break the continuous symmetry down to a discrete subgroup, thereby leading to the formation of cosmologically-problematic string-wall networks. 

The conventional way to annihilate DWs is by introducing an ad hoc bias term in the scalar potential~\cite{Gelmini:1988sf, Larsson:1996sp, Chang:1998tb, Vilenkin:2000jqa, Jana:2025vyb,Ai:2026zpy}. In this context, an alternative possibility has been explored more recently in the case of $Z_2$ DWs generated by a real scalar field, where a small bare fermion mass term is introduced in the Lagrangian, which can radiatively generate the required bias~\cite{Zhang:2023nrs,Zeng:2025zjp,Borah:2025bfa,Borah:2026kfo,Bhandari:2026ujy}\footnote{Ref.~\cite{Zhang:2023gfu} proposed an interesting alternative mechanism in which effective operators induce linear couplings between the Peccei-Quinn field and Standard Model particles, which ultimately generate a thermal potential that dynamically drives the axion field toward a universal value at high temperatures, thereby preventing DW formation in postinflationary axion models with discrete symmetry.}.


Here, we extend the radiatively induced bias mechanism previously studied in
simple $Z_2$ wall scenarios to the more general and nontrivial setting of
wall-string networks arising from the breaking of a global $U(1)$ symmetry
to a discrete subgroup $Z_n$ by gravitational effects. Unlike the $Z_2$ case, where an
isolated domain wall is present, or the $Z_3$ case, where the walls are necessarily adjacent, a generic $Z_n$ theory can contain
both adjacent and non-adjacent walls. This leads to a considerably complex
network structure and annihilation dynamics. In particular, we show that
the structure of the radiatively generated bias and the corresponding wall
annihilation depends systematically on whether $n$ is even or odd: a real
bare fermion mass is relevant for even $n$, whereas a complex bare mass,
with a nontrivial phase, is required for odd $n$. We establish these features
analytically for generic $n$, rather than restricting the analysis to a
particular discrete symmetry.

Moreover, for odd $n$, choosing a real bare
mass leads to a residual configuration in which a single wall is attached
to a string. Although this resembles the familiar single-wall-string
configuration, its stability is qualitatively different from
the conventional unstable configuration. This gives rise to an additional
stage in the evolution of the network and makes the annihilation dynamics richer. The sequential annihilation of non-adjacent walls can
further generate dynamically adjacent configurations before their subsequent
annihilation, providing a new microscopic picture of how the network evolves
toward the true vacuum. A schematic illustration of this DW annihilation mechanism in a string-wall system with multiple attached DWs is shown in Fig.~\ref{fig:BLscale}.

\vspace{0.1in}

\textbf{\emph{Formation of Domain Walls in a global $U(1)$ framework}.--}  As a representative example, we consider the well known global $U(1)_L$ majoron framework~\cite{Chikashige:1980ui}, in which the SM is extended by three singlet RHNs and a complex scalar $\Phi$ charged under the global $U(1)_L$ symmetry. When this symmetry is spontaneously broken, the RHNs acquire Majorana masses, while the light neutrino masses can be generated via the type-I seesaw mechanism~\cite{Minkowski:1977sc, Yanagida:1980xy, Gell-Mann:1979ijt, Gell-Mann:1979vob, Glashow:1979nm, Mohapatra:1979ia, Schechter:1980gr, Georgi:1981pg}.Here, $L$ denotes the lepton number.

Although we illustrate our discussion within this framework, our results apply more generally to all theories in which a global $U(1)$ symmetry is spontaneously broken by a complex scalar field that has Yukawa interactions with fermions.

As mentioned earlier, an important aspect of such scenarios is that global symmetries are generally expected to be violated by quantum gravitational effects near the Planck scale~\cite{Abbott:1989jw,Preskill:1992tc}. For instance, such explicit breaking may arise due to the nucleation of wormholes. Following the wormhole solution presented in Ref.~\cite{Abbott:1989jw}, for a field $\Phi$ carrying a nonzero charge under the global symmetry, integrating out wormhole configurations induces a low-energy effective term in the Lagrangian of the form
\begin{equation}
    \mathcal{L}_{\text{QG}} = \sum_{n=1}^{\infty} \mathcal{Y}_n g_n \, \Phi^n  +  \text{h.c.}\, ,
        \label{Eq.L_QG}
\end{equation}
where $\mathcal{Y}_n$ are complex parameters associated with the choice of vacuum state of the model and $n$ denotes the mass dimension of the operator. While, the effective couplings $g_n$ take the approximate form
\begin{equation}
    g_n \approx \left( \frac{\lambda^{(n-4)/3} n^{(n-16)/3}}{M_{Pl}^{(n-4)}} \right)
    e^{-\lambda^{1/3} n^{4/3}}
    \equiv \frac{\tilde{g}_{n}}{M_{Pl}^{(n-4)}} \, .
\end{equation}

Since the effective couplings $\tilde{g}_{n}$ are exponentially small due to the combined gravitational and scalar Euclidean action in the wormhole background, they are expected to be extremely small. For operator dimensions in the range $n=5$-$20$ and for $\lambda=1$, one typically finds $\tilde{g}_{n} \simeq 10^{-(n+2)}$, where $\lambda$ being the self coupling of the scalar $\Phi$~\cite{Abbott:1989jw}.

In principle, as shown in Eq.~\eqref{Eq.L_QG}, one can introduce operators starting from $n=1$ up to infinity. However, allowing $n\leq4$ operators with $\mathcal{O}(1)$ values of $\tilde g_n$ leads to hard breaking of the global $U(1)$ operator at $M_{Pl}$ itself and therefore such scenarios are not particularly interesting from a phenomenological perspective. For this reason, in our analysis we retain only operators with $n>4$. Lower dimensional operators with $n \leq 4$ can be forbidden in symmetry-driven constructions, for example by introducing an additional gauged $U(1)$ symmetry and another complex scalar, as discussed in Refs.~\cite{Rothstein:1992rh,Greljo:2025suh}. Nevertheless, the cosmology would become much more complicated~\cite{Ghosh:2025cxp}. Therefore, we do not consider those scenarios here and leave them for a future study. Furthermore, we also do not fix the value of $n$ to $n=5$. Instead, we keep it generic. Therefore, the leading-order gravity-motivated operator can be written as
\begin{equation}
    \mathcal{L}_{\text{lead}}
    \sim
    \frac{\tilde g_n}{M_{Pl}^{\, (n-4)}}
    \Big(
    \Phi^n
    +
    (\Phi^*)^n
    \Big).
    \label{qg}
\end{equation}

This operator explicitly breaks the global $U(1)_L$ symmetry and ultimately gives rise to the formation of a string-wall network. Here, we take $\tilde g_n$ to be a real parameter with $\tilde{g}_{n} = 10^{-(n+2)}$. The effects of subleading $(n+1)$-dimensional gravity-motivated operators are additionally suppressed by powers of $M_{\rm Pl}$ and are therefore neglected in our analysis.

Since $\Phi$ is a complex scalar field, after acquiring a vev, it can be decomposed in polar form as
\begin{align}
    \Phi
    =
    \frac{v+\rho}{\sqrt{2}}
    e^{i\chi/v},
    \label{polar_decom}
\end{align}
where $v$, $\rho$, and $\chi$ denote the vev of $\Phi$, the CP-even radial scalar, and the Goldstone boson (majoron) of the theory, respectively. Substituting Eq.~\eqref{polar_decom} into Eq.~\eqref{qg}, one finds that, at low energies after integrating out the heavy radial scalar $\rho$, the majoron field acquires the effective potential
\begin{align}
    V_{\text{majoron}}
    \sim
    -
    \frac{\tilde g_n}{2^{\,\frac{n}{2}-1}M_{Pl}^{\,n-4}}
    v^n
    \cos\!\left(\frac{n\chi}{v}\right).
    \label{majoron_pot}
\end{align}
The minima of the potential are determined by the condition $\cos\!\left(\frac{n\chi}{v}\right)=1$, implying
\[
\chi \big|_{\rm min}=\frac{2\pi kv}{n},
\qquad
k=0,1,\dots,n-1.
\]
Therefore, following Eq.~\eqref{polar_decom} and after integrating out the heavy scalar $\rho$, the corresponding $k$-dependent discrete vacua can be parametrized as~\cite{Preskill:1992ck} 
\begin{align}
    \langle\Phi\rangle_k
    =
    \frac{v}{\sqrt2}
    e^{i2\pi k/n}.
    \label{phi_vev_k}
\end{align}

The Eq.~\eqref{majoron_pot} has the form of standard Sine-Gordon potential, for which the majoron admits a DW solution of the form~\cite{Vilenkin:2000jqa}
\begin{align}
    \chi(x)
    =\frac{2\pi kv}{n}+
    \frac{4v}{n}
    \tan^{-1}\!\left(e^{m_\chi x}\right).
    \label{chi_x}
\end{align}

Here, $m_\chi$ denotes the majoron mass and can be approximated as
\begin{align}
m_\chi^2
\simeq
\frac{\tilde g_n n^2}{2^{\,\frac{n}{2}-1}}
\frac{v^{n-2}}{M_{\rm Pl}^{\,n-4}}.
\end{align}
The non-zero $m_\chi$ arises solely from the operator given in Eq.~\eqref{qg}, thereby rendering the majoron a pseudo-Nambu-Goldstone boson (pNGB).

During the cosmological evolution, a network of global cosmic strings (GCSs) first forms at a temperature $\sim v$. As the temperature decreases further to $T_f$, satisfying the condition $m_\chi \simeq 3H(T_f)$, with $H(T_f)$ denoting the Hubble parameter at the domain wall formation epoch $T_f$~\cite{Rothstein:1992rh}, the discrete vacuum structure becomes cosmologically relevant and domain walls form between each adjacent domains. As a consequence, each string becomes attached to $n$ domain walls. Furthermore, because of the cosine dependence of $V_{\rm majoron}$, the adjacent walls meet symmetrically around the string, as illustrated in Fig.~\ref{fig:BLscale}, thereby forming an energetically stable string-wall network configuration.
Furthermore, the surface tension of such majoron domain walls can be approximated as $\sigma_w \sim {8m_\chi v^2}/{n^2}$~\cite{Preskill:1992ck,Wu:2022tpe,Blasi:2023sej}. 

In the next section, we discuss how such DWs arising in the $n>4$ scenarios can be annihilated dynamically through the introduction of a tiny bare RHN mass term in the Lagrangian.

\vspace{0.1in}

\textbf{\emph{A  Mechanism for Domain Wall Annihilation in Global $U(1)$ Models}.--}
Once formed, these domain walls evolve under the influence of their surface tension. Numerical simulations suggest that the wall energy density scales as $\rho_w\sim a^{-r}$ with $1<r<2$, where $a$ denotes the scale factor of the Universe~\cite{Vilenkin:2000jqa}. Since this redshifts more slowly than both matter ($\sim a^{-3}$) and radiation ($\sim a^{-4}$), long-lived domain walls eventually come to dominate the energy density of the Universe, thereby substantially affecting the standard cosmological evolution and the process of structure formation, both of which are tightly constrained by BBN and other cosmological observations.

One of the most widely studied mechanisms for annihilating these walls is the introduction of a bias term in the scalar potential, which renders the DWs topologically unstable and cosmologically, once the pressure induced by the bias exceeds the wall energy density, they immediately start annihilating~\cite{Vilenkin:2000jqa, Berbig:2025nrt}.

However, in this section, rather than introducing such an ad hoc bias term in the scalar potential, we demonstrate a dynamical mechanism for the annihilation of these walls in the present majoron framework through the inclusion of a tiny bare RHN mass.

Motivated by the expectation that the global $U(1)_L$ symmetry is not exact in the presence of  gravitational effects, the inclusion of bare RHN mass terms in the Lagrangian is therefore not phenomenologically implausible~\cite{Giddings:1988cx,Heidenreich:2020pkc,Reece:2023czb}. In particular, in the context of wormholes and baby-universe formation, Ref.~\cite{Giddings:1988cx} argued that  gravitational effects generically induce all possible operators in the effective action that violate global symmetries. Therefore, it seems reasonable for global $U(1)_L$-breaking bare RHN mass terms to be present in the low-energy theory. In the following, we investigate how such bare RHN masses can dynamically trigger the annihilation of the string-wall network in this framework.

To illustrate this mechanism, let us first consider a bare Majorana mass term for a RHN field $N$ as,
\[
\mathcal{L}_{\rm bare}
=
\frac{1}{2}M_N \bar N^C N
+
\text{h.c.},
\]
which explicitly breaks $U(1)_L$ by two units. For simplicity, we consider a non-zero bare mass term for only a single RHN flavor and neglect all flavor mixing effects. From this point onward, the term RHN will refer exclusively to this particular flavor which has bare mass. Therefore, after the spontaneous breaking of the global $U(1)_L$ symmetry, the RHN Majorana mass term can be written as
\begin{align}
    \mathcal{L}_{M}
    =
    \frac{1}{2}\,
    \bar N^C
    \left(
    M_N
    +
    y\,\Phi
    \right)
    N
    +
    \text{h.c.}.
    \label{L_mr}
\end{align}

Here, $y$ denotes the Majorana Yukawa coupling between $N$ and  $\Phi$. Once $\Phi$ gets vev, employing Eq.~\eqref{phi_vev_k} we can parametrize RHN Majorana mass in the $k$-th domain as
\begin{align}
M_{R,k}
=
M_N
+
\frac{yv}{\sqrt{2}}
e^{i2\pi k/n}.
\label{Eq:M_R_k}
\end{align}
Because of the presence of this $M_N$, the phase factor $e^{i2\pi k/n}$ cannot be absorbed in the $|M_{R,k}|^2$ term. As we discuss in the subsequent paragraph, this nontrivial phase dependence in $|M_{R,k}|^2$ plays a crucial role in generating the bias required to annihilate the string-wall network. Here, different domains corresponding to different values of $k$ possess physically distinct RHN masses.

Since the RHN couples to the scalar $\Phi$ and admits a bare mass, we find that the one-loop Coleman-Weinberg (CW) effective potential generated by this RHN can contribute to the bias potential responsible for DW annihilation, which can be expressed in terms of the discrete vacuum label $k$ as~\cite{Coleman:1973jx,Gelmini:2020bqg} 
\begin{align}
    V_{CW}(k)\Big|_{\text{bias}}
    =-\frac{g}{64\pi^{2}}|M_{R,k}|^{4}
    \left[\ln\!\left(\frac{|M_{R,k}|^{2}}{\mu^{2}}\right)-\frac{3}{2}\right].
    \label{cw1}
\end{align}
Here $g(=2)$ denotes the number of degrees of freedom (d.o.f.) of that RHN species and the Feynman diagrams contributing to $V_{CW}(\chi)\big|_{\text{bias}}$ are shown in Fig.~\ref{fig:fermionloop}. Furthermore, for our analysis, we set the renormalization scale to $\mu = v$. 

In the limit $M_N\ll yv$, and neglecting the small logarithmic contribution in Eq.~\eqref{cw1}, the leading $k$-dependent contribution arises from the term linear in $M_N$. Expanding $|M_{R,k}|^4$ to first order in $M_N$, one obtains
\begin{align}
|M_{R,k}|^4
\simeq
\frac{(yv)^4}{4}
+
{\sqrt2}\,{(yv)^3}\,
M_N
\cos\!\left(\frac{2\pi k}{n}\right).
\end{align}
The first term is independent of $k$ and therefore does not contribute to the vacuum bias. Consequently, the bias between two adjacent vacua coming from CW-corrections corresponding to $k_1$ and $k_2$ can be approximated as
\begin{align}
V_{CW}\Big|_{\text{bias},\,k_1,k_2}
\simeq
\frac{3\,g}{64\sqrt{2}\,\pi^2}\,
M_{N}(yv)^{3}
\Bigg[
\cos\!\left(\frac{2\pi k_1}{n}\right)
\notag\\
-
\cos\!\left(\frac{2\pi k_2}{n}\right)
\Bigg].
\label{cw2}
\end{align}
This structure is quite similar to the Planck-suppressed bias discussed in Ref.~\cite{Blasi:2022ayo,Blasi:2023sej}. We present this approximated expression primarily to illustrate the analytical structure of the CW-induced bias term, which will be useful for understanding the DW dynamics discussed in the subsequent paragraphs. We stress that the characteristic dependence of $V_{CW}\big|_{\text{bias},\,k_1,k_2}$ on $[\cos(2\pi k_1/n)-\cos(2\pi k_2/n)]$ remains intact even when the full CW potential of Eq.~\eqref{cw1} is employed without any approximation.

\begin{figure}[t]

\centering


\begin{tikzpicture}[scale=.8]

\begin{scope}

\draw[dashed] (-2,0) -- (-0.9,0) node[midway,above] {$\Phi$};
\draw[dashed] (0.9,0) -- (2,0) node[midway,above] {$\Phi$};

\draw[thick] (0,0) circle (0.9);

\draw[->,thick] (110:0.9) arc (110:140:0.9);
\draw[->,thick] (290:0.9) arc (290:320:0.9);

\node at (0:0) {$N$};

\end{scope}

\node at (3,0) {$+$};

\begin{scope}[xshift=6cm]

\draw[dashed] (-2,0) -- (-0.9,0) node[midway,above] {$\Phi$};
\draw[dashed] (0.9,0) -- (2,0) node[midway,above] {$\Phi$};
\draw[dashed] (0,0.9) -- (0,2) node[midway,right] {$\Phi$};
\draw[dashed] (0,-0.9) -- (0,-2) node[midway,right] {$\Phi$};

\draw[thick] (0,0) circle (0.9);

\draw[->,thick] (30:0.9) arc (30:60:0.9);
\draw[->,thick] (120:0.9) arc (120:150:0.9);
\draw[->,thick] (210:0.9) arc (210:240:0.9);
\draw[->,thick] (300:0.9) arc (300:330:0.9);

\node at (0:0) {$N$};

\end{scope}

\end{tikzpicture}

\vspace{0.3cm}


\begin{tikzpicture}[scale=0.8]

\node at (-3.5,0) {$+$};

\foreach \a in {0,60,120,180,240,300}
{
\draw[dashed] ({0.9*cos(\a)},{0.9*sin(\a)})
-- ({2*cos(\a)},{2*sin(\a)})
node[midway,sloped,above] {$\Phi$};
}

\draw[thick] (0,0) circle (0.9);

\foreach \a in {30,90,150,210,270,330}
{
\draw[->,thick] (\a:0.9) arc (\a:\a+20:0.9);
}

\node at (0:0) {$N$};

\node at (3.5,0) {$+\cdots$};

\end{tikzpicture}

\caption{1-loop Feynman diagrams contributing to the Coleman-Weinberg effective potential for the scalar field $\Phi$, generated through the $k$-dependent RHN mass $M_{R,k}$.}

\label{fig:fermionloop}

\end{figure}


Additionally, there can also be a one-loop thermal correction to the bias potential arising from the same RHN field, which can be written as
\begin{align}
V_{J_{F}}\!\left(\frac{|M_{R,k}|^{2}}{T^{2}}\right)\Bigg|_{\text{bias}}
= -\frac{gT^{4}}{2\pi^{2}}
   \int_{0}^{\infty} dx\, x^{2} \notag \\
\times
   \log\!\left[
   1 + \exp\!\left(
   -\sqrt{x^{2} + \frac{|M_{R,k}|^{2}}{T^{2}}}
   \right)
   \right].
   \label{V_jf}
\end{align}

\begin{figure*}[t]
\centering
\includegraphics[width=0.48\textwidth]{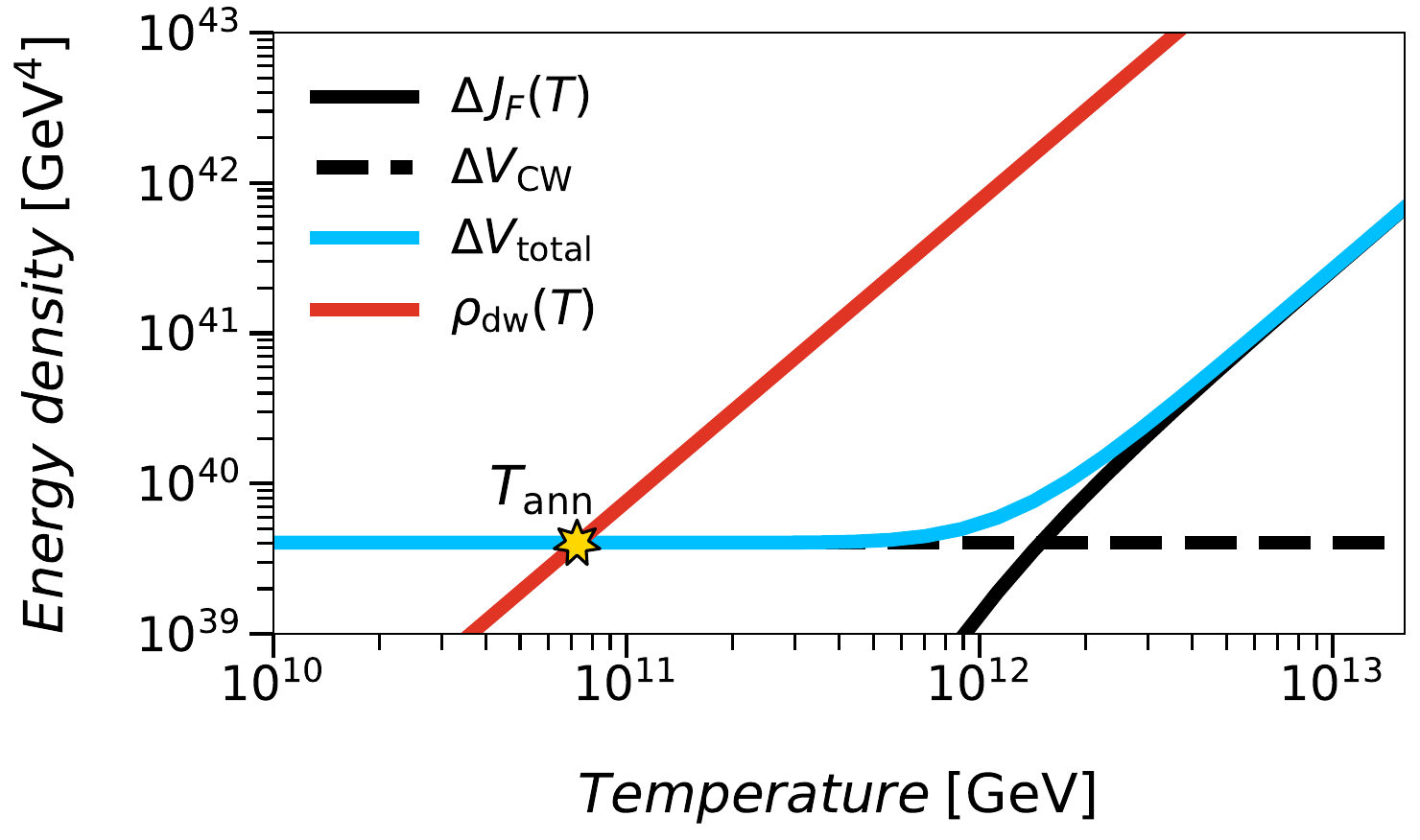}
\hfill
\includegraphics[width=0.48\textwidth]{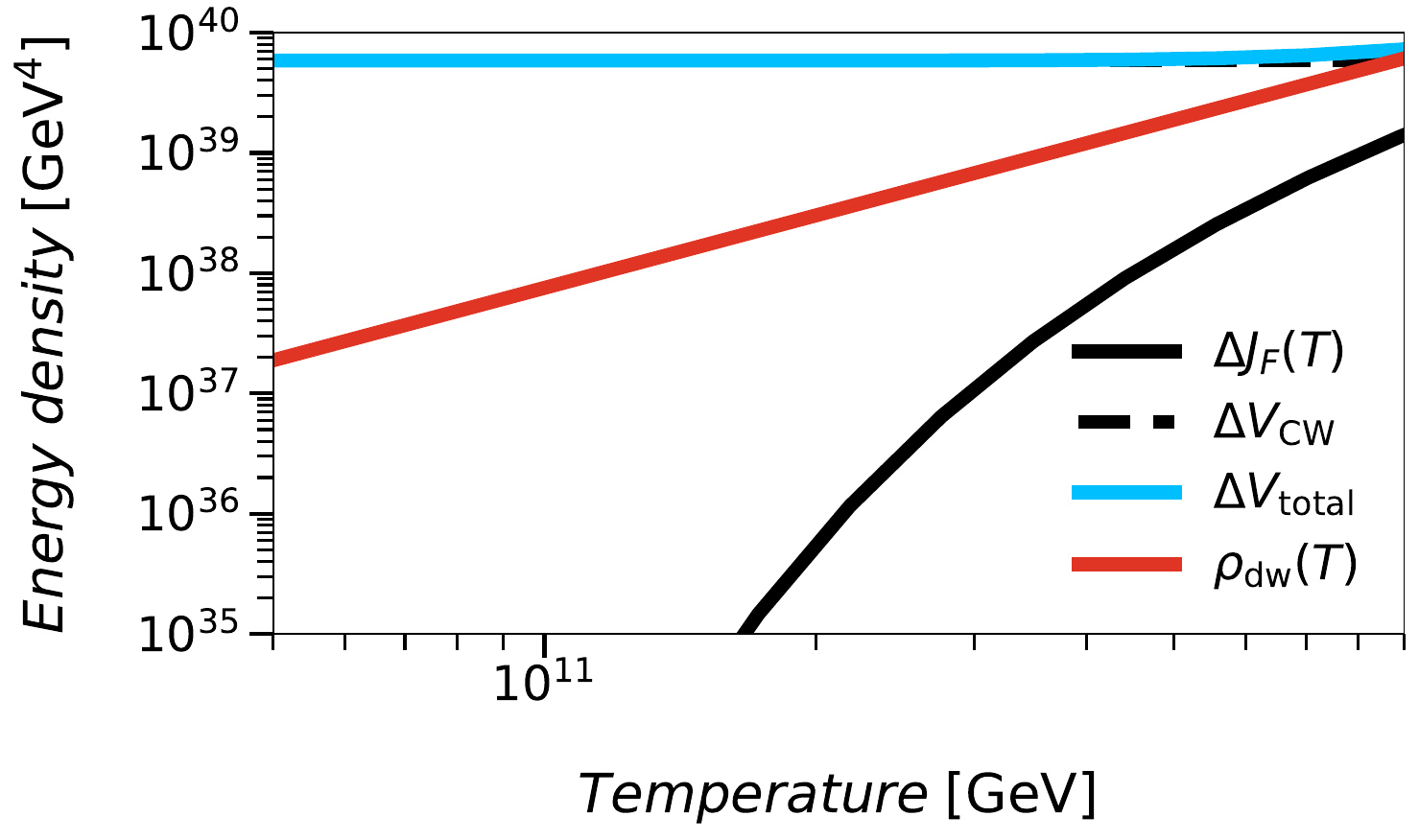}
\caption{Evolution of the domain wall and bias energy densities for the two benchmark points listed in Table~\ref{tab:benchmark}. 
Left panel: BP1 ($n=5$), where the domain walls form at $T_f=2\times 10^{13}$ GeV and subsequently annihilate at 
$T_{\rm ann}=7.25 \times 10^{10}\,\mathrm{GeV}$ for a benchmark value of $\delta=2\pi/6$. 
Right panel: BP2 ($n=6$), for which the bias term dominates over wall energy density even at the DW formation time, thereby preventing the domain wall formation from the beginning.
}

\label{fig:placeholder}
\end{figure*}


 Therefore, we can write the $k$-dependent effective total bias potential associated with the $\Phi$ field at a nonzero temperature $T$, including one-loop corrections as
\begin{align}
    V_{\text{bias, total}}(k,T)
    &=
    V_o(\Phi)\Big|_{\Phi=\langle\Phi\rangle_k}+V_{\text{majoron}}(\chi)\Big|_{\chi={2\pi kv}/{n}}
    \notag \\
    &\quad
    +V_{CW}(k)\Big|_{\text{bias}}
    +V_{J_{F}}\!\left(\frac{|M_{R,k}|^{2}}{T^{2}}\right)\Bigg|_{\text{bias}}.
    \label{V_tot}
\end{align}

The first term corresponds to the tree-level scalar potential of the field $\Phi$, $V_0(\Phi)=\lambda(\Phi^*\Phi-\langle\Phi\rangle_k^*\langle\Phi\rangle_k)^2$, while the second term exactly mimics Eq.~\eqref{qg} in the potential form. Here we assume that the pNGB majoron $\chi$ and the RHNs are in thermal equilibrium with a bath at temperature $T$, which is necessary for obtaining a nonzero contribution from $V_{J_F}$. Nevertheless, for the benchmark points listed in Tab.~\ref{tab:benchmark}, we find that the annihilation of the DWs is predominantly driven by the temperature-independent CW corrections, while the thermal contribution remains subdominant, as illustrated in Fig.~\ref{fig:placeholder}. Therefore, although both $V_{CW}$ and $V_{J_F}$ 1-loop contributions are included in the full expression below, the dominant source of the vacuum bias arises from the CW term. The total bias between two vacua labeled by $k_1$ and $k_2$ can thus be expressed as
\begin{align}
    \Delta V\big|_{k_1, k_2}
    =\big|V_{\text{bias, total}}\!\left(k_1,T\right) 
    -V_{\text{bias, total}}\!\left(k_2,T\right)\big|.
    \label{delta_v}
\end{align}

By construction, the first term of Eq.~\eqref{V_tot} is $U(1)$ invariant, and therefore 
$(V_0(\Phi)|_{\Phi=\langle\Phi\rangle_{k_1}}
-
V_0(\Phi)|_{\Phi=\langle\Phi\rangle_{k_2}})=0$. Furthermore, the contribution arising from the second term in Eq.~\eqref{V_tot} also exactly cancels, i.e.

$(V_{\text{majoron}}|_{k_1}-V_{\text{majoron}}|_{k_2})=0$, since the argument of the cosine in Eq.~\eqref{majoron_pot} is given by $n\chi/v$, while the vacua are located at $\chi=2\pi kv/n$. Consequently, $\cos(n\chi/v)=\cos(2\pi k)=1$ for any integer $k$, implying that $V_{\text{majoron}}=\text{constant}$ for all vacua. This conclusion remains unchanged even if a phase factor $e^{i\alpha}$ is included in the Planck-suppressed operator of Eq.~\eqref{qg}, corresponding to a complex $\tilde g_n$, since $\cos(2\pi k+\alpha)=\cos\alpha=\text{constant}$ for all $k\in\mathbb{Z}$. In contrast, the CW and thermal corrections in Eq.~\eqref{V_tot} generate nonzero contributions to $\Delta V$ due to the nontrivial structure of the potentials given in Eq.~\eqref{cw2} and Eq.~\eqref{V_jf}. An exception arises when two vacua labeled by $k_1$ and $k_2$ satisfy $k_1+k_2=n$,\footnote{In general, the condition for vanishing bias between two vacua, $k_1$ and $k_2$, is $k_1+k_2=mn$, where $m\in\{1,2,3,\ldots\}$, as follows from equating the corresponding cosine terms (since both $k_1$ and $k_2$ are positive integers, $m$ cannot be zero). Since the distinct vacua are labeled by $k=0,1,\ldots,n-1$, the maximum possible value of $k_1+k_2$ is $2n-3$, attained for $(k_1,k_2)=(n-1,n-2)$. Therefore, $k_1+k_2<2n$, which implies that the only possible positive integer value of $m$ is $m=1$. Hence, for two distinct vacua, the most general condition for vanishing bias is $k_1+k_2=n$.} In this case,
\[
\cos\!\left(\frac{2\pi k_2}{n}\right)
=
\cos\!\left(2\pi-\frac{2\pi k_1}{n}\right)
=
\cos\!\left(\frac{2\pi k_1}{n}\right),
\]
which immediately causes the CW contribution in Eq.~\eqref{cw2} to vanish. The same condition also implies that the thermal contributions become identical, since the thermal potential in Eq.~\eqref{V_jf} depends only on $|M_{R,k}|^2$, where
\[
|M_{R,k}|^2
=
M_N^2+\frac{(yv)^2}{2}
+\sqrt2\,M_Nyv
\cos\!\left(\frac{2\pi k}{n}\right).
\]
Using the above cosine relation, one finds $|M_{R,k_1}|^2=|M_{R,k_2}|^2$ and therefore the thermal contribution also yields $\Delta V=0$. This demonstrates the crucial role of the bare mass term, since in its absence one always obtains $\Delta V=0$ for all vacua.\\

\begin{table*}[t]
\centering
\renewcommand{\arraystretch}{1.3}
\begin{ruledtabular}
\begin{tabular}{ccccccccc}
Benchmark Point & $n$ & $\tilde{g}_n=10^{-(n+2)}$ & $M_N$ (GeV)& $\delta$ & $y$ & $v$ (GeV) & $T_f$ (GeV) & $T_{\rm ann}$ (GeV) \\
\hline
BP1 & 5 & $10^{-7}$ & $10^{4}$ & $\frac{2\pi}{6}$ & $0.01$ & $2\times10^{14}$ & $2 \times 10^{13}$ & $7.25\times10^{10}$ \\
BP2 & 6 & $10^{-8}$ & $10^{4}$ & 0 & $0.01$ & $2\times10^{14}$ & $9 \times 10^{11}$ & $T_f$ \\
\end{tabular}
\end{ruledtabular}
\caption{Benchmark points corresponding to $n=5$ and $n=6$. For BP1 ($n=5$), the annihilation temperature is $T_{\rm ann}=7.25 \times 10^{10}\,\mathrm{GeV}$, as shown in the left panel of Fig.~\ref{fig:placeholder}. BP2 corresponds to $n=6$, for which $T_{\rm ann}$ is not well defined, since the bias term already dominates over the wall energy density at the time of wall formation, $T_f$, as shown in the right panel of Fig.~\ref{fig:placeholder}.}
\label{tab:benchmark}
\end{table*}
Since a domain wall cannot annihilate when the bias between the two vacua that it interpolates vanishes, the vacuum structure of the explicit symmetry-breaking potential plays the central role in determining the final fate of the string-wall network. Although different microscopic annihilation histories are possible depending on which wall experiences the largest bias at a given stage, these details do not affect the final outcome.

For even values of $n$, the cosine potential possesses a unique global minimum. Consequently, irrespective of the annihilation sequence, every domain wall eventually experiences a non-vanishing pressure toward this true vacuum. The network therefore continues to collapse until only the cosmic string (in the core of the system) remains. The $n=6$ example discussed in the End Matter represents one possible annihilation sequence. However, many other microscopic histories are equally possible, and all lead to the same final configuration because the true vacuum is unique.

The situation is qualitatively different for odd values of $n$. In this case, the minimum of the cosine potential is not unique. Instead, it is shared by two neighbouring vacua satisfying the condition $k_1+k_2=n$, which are exactly degenerate. Consequently, irrespective of the annihilation history, the network ultimately reduce to a configuration consisting of a cosmic string attached to a single domain wall separating these two vacua. Since the vacuum-energy difference across this wall vanishes, it experiences no pressure to collapse and therefore remains stable. For example, in the $n=5$ case, the adjacent vacua $(k_1,k_2)=(2,3)$ satisfy this condition. Thus, unlike the even-$n$ case, the cosmological domain wall problem persists for odd values of $n$.

The above discussion immediately shows that introducing only a real bare Majorana mass $M_N$ is insufficient to resolve the problem. Although $M_N$ explicitly breaks the global $U(1)$ symmetry and generates a non-vanishing bias between most pairs of vacua, it cannot lift the degeneracy of the pair satisfying $k_1+k_2=n$. Therefore, an additional source of explicit symmetry breaking is required.

A simple way to achieve this is to allow the RHN bare Majorana mass to carry a constant complex phase,
\begin{equation}
M_N \rightarrow M_N e^{-i\delta},
\end{equation}
with $0<\delta<2\pi$. The condition for vanishing bias is then modified to
\begin{equation}
k_1+k_2+\frac{n\delta}{\pi}=ln,
\label{Eq.ln}
\end{equation}
where $l=1,2,3,\ldots$.
\footnote{Since $k_1$ and $k_2$ take positive integer values and $0<\delta<2\pi$, $l$ cannot be zero.}
Consequently, for a generic non-zero value of $\delta$, the pair of vacua that previously satisfied $k_1+k_2=n$ is no longer degenerate, thereby generating a non-vanishing bias across the remaining wall. Consequently, the last surviving wall also annihilates, thereby resolving the cosmological domain-wall problem for odd values of $n$.

A related role of a phase \(\delta\) was discussed in Ref.~\cite{Wu:2022stu} in the context of \(Z_3\) domain walls, where the phase is introduced in the explicit \(Z_3\)-breaking scalar potential at tree level, generating a bias among the vacua, which are all mutually adjacent in the \(Z_3\) case. By contrast, in our case, \(\delta\) enters through the bare Majorana mass, \(M_N e^{-i\delta}\), and induces the bias radiatively through the fermion-scalar Yukawa interaction (with coupling strength \(y\)); hence, the bias vanishes as \(y\to0\). Moreover, while Ref.~\cite{Wu:2022stu} considers only \(Z_3\) walls, we study the general \(U(1)\to Z_n\) string-wall system, which includes initially non-neighbouring vacua that can become dynamically adjacent, and derive the phase-dependent condition for vanishing bias and the possible residual degeneracies for general \(n\), which we discuss in the subsequent paragraphs.\footnote{Another interesting study of the same ordinary $Z_3$ domain-wall annihilation was presented in Ref.~\cite{Li:2025gld}, where the authors investigated how a preferred bias direction influences the annihilation dynamics.}

From Eq.~(\ref{Eq.ln}), it is evident that the only subtlety in eliminating the vanishing-bias configurations arises when the phase takes the special form
\begin{equation}
\delta=p\frac{\pi}{n},
\end{equation}
where $p$ is an integer. In this case, the modified condition becomes
\begin{equation}
k_1+k_2+p=ln.
\end{equation}

For $l=1$, the condition reduces to
\begin{equation}
k_1+k_2+p=n.
\end{equation}
Although the original degenerate pair satisfying $k_1+k_2=n$ is lifted, another pair of vacua may satisfy this condition and become degenerate. For example, for $n=5$ and $p=2$, corresponding to $\delta=2\pi/5$, the original pair $(2,3)$ is lifted, while the pair $(1,2)$ becomes degenerate. Similarly, for $p=4$, the pair $(0,1)$ becomes degenerate. Thus, for $0\leq p<n$, accidental degeneracies may remain. For $l=1$, these degeneracies can be avoided by choosing $p\geq n$, since $k_1+k_2\leq 2n-3$ and hence the condition $k_1+k_2+p=n$ cannot be satisfied.

For $l>1$, however, the condition $p\geq n$ is no longer sufficient to eliminate all accidental degeneracies. For instance, taking $n=5$, $l=3$, and $p=8>n$, the pair $(k_1,k_2)=(3,4)$ satisfies
\begin{equation}
k_1+k_2+p=3+4+8=15=3n,
\end{equation}

and therefore remains degenerate. Hence, a choice of $p$ that removes the degeneracy for $l=1$ may still leave accidental degeneracies for higher values of $l$.

A generic phase that is not an integer multiple of $\pi/n$ provides a robust solution. In this case, $n\delta/\pi$ is non-integer, so the left-hand side of the modified condition is non-integer for any pair of vacua, whereas the right-hand side is always an integer. Consequently, no pair of vacua can satisfy the vanishing-bias condition, ensuring that every domain wall acquires a non-zero bias and that the complete string-wall network eventually annihilates.

For successful domain-wall annihilation, the bias energy density must exceed the wall energy density, namely $\Delta V(T)\gtrsim\rho_w(T)$. Following Ref.~\cite{Hiramatsu:2012sc}, we parametrize the wall energy density as $\rho_w(T)\simeq\sigma_wH(T)$. The annihilation temperature $T_{\rm ann}$ is determined from the condition $\Delta V(T_{\rm ann})\simeq\rho_w(T_{\rm ann})$, together with the consistency requirement $T_{\rm ann}>T_{\rm BBN}$.

Since different values of $k$ correspond to different values of $V_{\rm bias,total}$, different walls generally possess different annihilation temperatures. Nevertheless, these differences remain relatively small. For the $n=5$ benchmark, the spread in annihilation temperatures satisfies $\Delta T_{\rm ann}/T_{\rm ann}\sim\mathcal{O}(50\%)$, implying that all walls annihilate nearly simultaneously on cosmological timescales. We further find that this relative spread decreases for larger values of $n$.

Additionally, unlike conventional approaches in which explicit bias terms of the form $\Phi^m$ (with $m$ co-prime to $n$) are introduced by hand, our framework dynamically generates the required vacuum bias through radiative corrections originating from the bare RHN mass.

Another important consequence of the bare Majorana mass is that the majoron acquires both derivative and non-derivative couplings to the RHNs~\cite{Rothstein:1992rh,Ghosh:2025cxp}. These interactions can generate friction on the moving domain walls if the RHNs remain in thermal equilibrium. Although such friction may modify the approach to the scaling regime and consequently shift $T_{\rm ann}$, it does not alter the topology of the vacuum manifold and therefore has no impact on whether domain walls ultimately annihilate. Since our focus is on the existence and annihilation of the walls themselves, friction effects are neglected throughout this work.

For illustration, we consider two benchmark points corresponding to $n=5$ and $n=6$. For BP1, we choose $\{M_N=10^4~{\rm GeV},\,y=0.01,\,\tilde g_5=10^{-7},\,v=2\times10^{14}~{\rm GeV}\}$, yielding $T_f\simeq2\times10^{13}$ GeV and $T_{\rm ann}\simeq 7.25\times10^{10}$ GeV. For BP2, corresponding to $n=6$, we take $\tilde g_6=10^{-8}$ while keeping the remaining benchmark parameters unchanged. In this case, the bias already exceeds the wall energy density at the formation epoch, $T_f\simeq9\times10^{11}$ GeV, preventing the domain walls from surviving after formation. We further find that this behaviour persists for all higher even-dimensional operators with $n>6$ for the same benchmark parameters.


\textbf{\emph{Conclusions}.--}
In this work, we have proposed a mechanism for triggering the annihilation of cosmological domain walls in theories with global $U(1)$ symmetries. Quantum-gravity-induced higher-dimensional operators explicitly break the continuous symmetry to a discrete subgroup, leading to the formation of string-wall networks. We have shown that radiative corrections induced by a tiny bare Majorana mass of the RHNs generate the vacuum bias required for domain-wall annihilation. An important outcome of our analysis is the qualitative distinction between even- and odd-dimensional operators. For even values of $n$, the explicit symmetry-breaking potential possesses a unique true vacuum, ensuring that the entire wall network annihilates irrespective of the microscopic annihilation history. In contrast, for odd values of $n$, the potential admits two degenerate neighbouring minima, allowing one stable wall to survive. We have shown that this remaining degeneracy can be lifted by introducing a constant complex phase in the RHN bare mass term, thereby generating a non-vanishing bias across the final wall and completely resolving the cosmological domain wall problem. As a concrete realization, we have illustrated the mechanism within the global $U(1)_L$ majoron framework.

\vspace{0.1in}
\begin{acknowledgments}
{\textbf {\textit {Acknowledgments.--}}}
The authors would like to thank Subhojit Roy, Anirban Basu, Santosh Kumar Rai, and Sudip Jana for useful comments and discussions. The authors are also grateful to acknowledge the support of the Harish-Chandra Research Institute and Homi Bhabha National Institute (HBNI), Mumbai.
\end{acknowledgments}

\appendix

\section*{End Matter}
\textbf{\emph{Detailed Discussion of String-Wall Network Annihilation for $n\geq5$.--}}

For the illustrative case of $n=5$, the bias responsible for domain-wall annihilation is predominantly generated by the Coleman-Weinberg contribution given in Eq.~\eqref{cw2}. Consequently, the biases between adjacent vacua are proportional to
\begin{align}
\Delta V_{01}
&\propto
\left|
\cos(0)-\cos\!\left(\frac{2\pi}{5}\right)
\right|
\simeq0.69,\\
\Delta V_{12}
&\propto
\left|
\cos\!\left(\frac{2\pi}{5}\right)
-
\cos\!\left(\frac{4\pi}{5}\right)
\right|
\simeq1.12,\\
\Delta V_{23}
&\propto
\left|
\cos\!\left(\frac{4\pi}{5}\right)
-
\cos\!\left(\frac{6\pi}{5}\right)
\right|
=0,\\
\Delta V_{34}
&\propto
\left|
\cos\!\left(\frac{6\pi}{5}\right)
-
\cos\!\left(\frac{8\pi}{5}\right)
\right|
\simeq1.12,
\end{align}
\begin{align}
\Delta V_{40}
&\propto
\left|
\cos\!\left(\frac{8\pi}{5}\right)
-
\cos(0)
\right|
\simeq0.69.
\end{align}

Thus, among all adjacent vacua, only the wall interpolating between $(k_1,k_2)=(2,3)$ has vanishing bias from the moment the domain-wall network forms. Since the walls connecting the pairs $(1,2)$ and $(3,4)$ possess the largest bias, they annihilate first. During this process, the vacua $k=2$ and $k=3$ are energetically favoured over $k=1$ and $k=4$, respectively. Consequently, after these annihilations, the remaining neighbouring domains become $(0,2)$, $(2,3)$ and $(3,0)$.

The domains $(0,2)$ and $(3,0)$ are now dynamically adjacent, and their corresponding biases are
\begin{align}
\Delta V_{02}
&\propto
\left|
\cos(0)
-
\cos\!\left(\frac{4\pi}{5}\right)
\right|
\simeq1.81,\\
\Delta V_{30}
&\propto
\left|
\cos\!\left(\frac{6\pi}{5}\right)
-
\cos(0)
\right|
\simeq1.81.
\end{align}

These walls therefore annihilate immediately, leaving only the unbiased wall separating the degenerate vacua $(2,3)$. Hence, although several walls initially coexist in the network, the sequential annihilation dynamics eventually reduces the system to a configuration consisting of a cosmic string attached to a single domain wall. Unlike the even-$n$ case, however, this remaining wall is stable because it separates two exactly degenerate vacua. Therefore, the annihilation mechanism cannot proceed any further. As discussed in the main text, this remaining degeneracy can be lifted only by introducing a constant complex phase in the bare RHN mass,
\begin{equation}
M_N\rightarrow M_Ne^{-i\delta},
\end{equation}
which generates a non-vanishing bias across the final wall and allows it to annihilate.

For the illustrative even-dimensional example $n=6$, the adjacent-wall biases become
\begin{align}
\Delta V_{01}&\propto0.5,\\
\Delta V_{12}&\propto1,\\
\Delta V_{23}&\propto0.5,\\
\Delta V_{34}&\propto0.5,\\
\Delta V_{45}&\propto1,\\
\Delta V_{50}&\propto0.5.
\end{align}

Unlike the $n=5$ case, every adjacent wall possesses a non-vanishing bias from the moment the network forms. One possible annihilation sequence proceeds as follows. Since the walls connecting $(1,2)$ and $(4,5)$ possess the largest bias, they annihilate first, leaving the vacua $k=2$ and $k=4$ energetically favoured over $k=1$ and $k=5$. Consequently, the remaining neighbouring domains become $(0,2)$, $(0,4)$, $(2,3)$ and $(3,4)$.

The biases among these remaining neighbouring domains are
\begin{align}
\Delta V_{02}&\propto1.5,\\
\Delta V_{40}&\propto1.5,\\
\Delta V_{23}&\propto0.5,\\
\Delta V_{34}&\propto0.5.
\end{align}

The walls connecting $(0,2)$ and $(0,4)$ therefore annihilate next, eliminating the false vacuum $k=0$. As a consequence, only the walls separating $(2,3)$ and $(3,4)$ remain. Since both of these walls also possess non-vanishing bias, they subsequently annihilate as well, driving the system completely into the unique true vacuum corresponding to $k=3$. This example illustrates the generic behaviour discussed in the main text: for even values of $n$, the explicit symmetry-breaking potential possesses a unique global minimum, and therefore the complete string-wall network annihilates without requiring any additional phase $\delta$. The bare RHN mass $M_N$ alone is sufficient to resolve the cosmological domain-wall problem.

For larger odd values of $n>5$, the qualitative picture remains unchanged. Although the detailed annihilation history may differ, there always exists one pair of degenerate vacua satisfying $k_1+k_2=n$. During the annihilation process, this pair need not remain adjacent at all times, and previously non-adjacent degenerate vacua may become dynamically adjacent after intermediate walls disappear. Nevertheless, irrespective of the microscopic annihilation history, the network ultimately reduces to a configuration consisting of a cosmic string attached to a single domain wall separating two degenerate vacua. Therefore, exactly as in the $n=5$ example, an additional complex phase $\delta$ is required to lift the remaining degeneracy and complete the annihilation.

For even values of $n>6$, the qualitative behaviour is likewise identical to the $n=6$ example. Although different microscopic annihilation sequences are possible and previously non-adjacent vacua may become dynamically adjacent during the evolution, the unique true vacuum ensures that every remaining wall always experiences a non-vanishing bias. Consequently, irrespective of the annihilation history, the complete string-wall network inevitably collapses without introducing any additional parameter beyond the bare RHN mass.
\bibliographystyle{apsrev4-1}
\bibliography{reference}

\begin{thebibliography}{49}%
\makeatletter
\providecommand \@ifxundefined [1]{%
 \@ifx{#1\undefined}
}%
\providecommand \@ifnum [1]{%
 \ifnum #1\expandafter \@firstoftwo
 \else \expandafter \@secondoftwo
 \fi
}%
\providecommand \@ifx [1]{%
 \ifx #1\expandafter \@firstoftwo
 \else \expandafter \@secondoftwo
 \fi
}%
\providecommand \natexlab [1]{#1}%
\providecommand \enquote  [1]{``#1''}%
\providecommand \bibnamefont  [1]{#1}%
\providecommand \bibfnamefont [1]{#1}%
\providecommand \citenamefont [1]{#1}%
\providecommand \href@noop [0]{\@secondoftwo}%
\providecommand \href [0]{\begingroup \@sanitize@url \@href}%
\providecommand \@href[1]{\@@startlink{#1}\@@href}%
\providecommand \@@href[1]{\endgroup#1\@@endlink}%
\providecommand \@sanitize@url [0]{\catcode `\\12\catcode `\$12\catcode
  `\&12\catcode `\#12\catcode `\^12\catcode `\_12\catcode `\%12\relax}%
\providecommand \@@startlink[1]{}%
\providecommand \@@endlink[0]{}%
\providecommand \url  [0]{\begingroup\@sanitize@url \@url }%
\providecommand \@url [1]{\endgroup\@href {#1}{\urlprefix }}%
\providecommand \urlprefix  [0]{URL }%
\providecommand \Eprint [0]{\href }%
\providecommand \doibase [0]{http://dx.doi.org/}%
\providecommand \selectlanguage [0]{\@gobble}%
\providecommand \bibinfo  [0]{\@secondoftwo}%
\providecommand \bibfield  [0]{\@secondoftwo}%
\providecommand \translation [1]{[#1]}%
\providecommand \BibitemOpen [0]{}%
\providecommand \bibitemStop [0]{}%
\providecommand \bibitemNoStop [0]{.\EOS\space}%
\providecommand \EOS [0]{\spacefactor3000\relax}%
\providecommand \BibitemShut  [1]{\csname bibitem#1\endcsname}%
\let\auto@bib@innerbib\@empty
\bibitem [{\citenamefont {Kibble}(1976)}]{Kibble:1976sj}%
  \BibitemOpen
  \bibfield  {author} {\bibinfo {author} {\bibfnamefont {T.~W.~B.}\
  \bibnamefont {Kibble}},\ }\href {\doibase 10.1088/0305-4470/9/8/029}
  {\bibfield  {journal} {\bibinfo  {journal} {J. Phys. A}\ }\textbf {\bibinfo
  {volume} {9}},\ \bibinfo {pages} {1387} (\bibinfo {year} {1976})}\BibitemShut
  {NoStop}%
\bibitem [{\citenamefont {Vilenkin}\ and\ \citenamefont
  {Everett}(1982)}]{Vilenkin:1982ks}%
  \BibitemOpen
  \bibfield  {author} {\bibinfo {author} {\bibfnamefont {A.}~\bibnamefont
  {Vilenkin}}\ and\ \bibinfo {author} {\bibfnamefont {A.~E.}\ \bibnamefont
  {Everett}},\ }\href {\doibase 10.1103/PhysRevLett.48.1867} {\bibfield
  {journal} {\bibinfo  {journal} {Phys. Rev. Lett.}\ }\textbf {\bibinfo
  {volume} {48}},\ \bibinfo {pages} {1867} (\bibinfo {year}
  {1982})}\BibitemShut {NoStop}%
\bibitem [{\citenamefont {Bogomolny}(1976)}]{Bogomolny:1975de}%
  \BibitemOpen
  \bibfield  {author} {\bibinfo {author} {\bibfnamefont {E.~B.}\ \bibnamefont
  {Bogomolny}},\ }\href@noop {} {\bibfield  {journal} {\bibinfo  {journal}
  {Sov. J. Nucl. Phys.}\ }\textbf {\bibinfo {volume} {24}},\ \bibinfo {pages}
  {449} (\bibinfo {year} {1976})}\BibitemShut {NoStop}%
\bibitem [{\citenamefont {'t~Hooft}(1974)}]{tHooft:1974kcl}%
  \BibitemOpen
  \bibfield  {author} {\bibinfo {author} {\bibfnamefont {G.}~\bibnamefont
  {'t~Hooft}},\ }\href {\doibase 10.1016/0550-3213(74)90486-6} {\bibfield
  {journal} {\bibinfo  {journal} {Nucl. Phys. B}\ }\textbf {\bibinfo {volume}
  {79}},\ \bibinfo {pages} {276} (\bibinfo {year} {1974})}\BibitemShut
  {NoStop}%
\bibitem [{\citenamefont {Polyakov}(1974)}]{Polyakov:1974ek}%
  \BibitemOpen
  \bibfield  {author} {\bibinfo {author} {\bibfnamefont {A.~M.}\ \bibnamefont
  {Polyakov}},\ }\href@noop {} {\bibfield  {journal} {\bibinfo  {journal} {JETP
  Lett.}\ }\textbf {\bibinfo {volume} {20}},\ \bibinfo {pages} {194} (\bibinfo
  {year} {1974})}\BibitemShut {NoStop}%
\bibitem [{\citenamefont {Nambu}(1977)}]{Nambu:1977ag}%
  \BibitemOpen
  \bibfield  {author} {\bibinfo {author} {\bibfnamefont {Y.}~\bibnamefont
  {Nambu}},\ }\href {\doibase 10.1016/0550-3213(77)90252-8} {\bibfield
  {journal} {\bibinfo  {journal} {Nucl. Phys. B}\ }\textbf {\bibinfo {volume}
  {130}},\ \bibinfo {pages} {505} (\bibinfo {year} {1977})}\BibitemShut
  {NoStop}%
\bibitem [{\citenamefont {Kibble}\ and\ \citenamefont
  {Vachaspati}(2015)}]{Kibble:2015twa}%
  \BibitemOpen
  \bibfield  {author} {\bibinfo {author} {\bibfnamefont {T.~W.~B.}\
  \bibnamefont {Kibble}}\ and\ \bibinfo {author} {\bibfnamefont
  {T.}~\bibnamefont {Vachaspati}},\ }\href {\doibase
  10.1088/0954-3899/42/9/094002} {\bibfield  {journal} {\bibinfo  {journal} {J.
  Phys. G}\ }\textbf {\bibinfo {volume} {42}},\ \bibinfo {pages} {094002}
  (\bibinfo {year} {2015})},\ \Eprint {http://arxiv.org/abs/1506.02022}
  {arXiv:1506.02022 [astro-ph.CO]} \BibitemShut {NoStop}%
\bibitem [{\citenamefont {Rothstein}\ \emph {et~al.}(1993)\citenamefont
  {Rothstein}, \citenamefont {Babu},\ and\ \citenamefont
  {Seckel}}]{Rothstein:1992rh}%
  \BibitemOpen
  \bibfield  {author} {\bibinfo {author} {\bibfnamefont {I.~Z.}\ \bibnamefont
  {Rothstein}}, \bibinfo {author} {\bibfnamefont {K.~S.}\ \bibnamefont {Babu}},
  \ and\ \bibinfo {author} {\bibfnamefont {D.}~\bibnamefont {Seckel}},\ }\href
  {\doibase 10.1016/0550-3213(93)90368-Y} {\bibfield  {journal} {\bibinfo
  {journal} {Nucl. Phys. B}\ }\textbf {\bibinfo {volume} {403}},\ \bibinfo
  {pages} {725} (\bibinfo {year} {1993})},\ \Eprint
  {http://arxiv.org/abs/hep-ph/9301213} {arXiv:hep-ph/9301213} \BibitemShut
  {NoStop}%
\bibitem [{\citenamefont {Turok}(1989)}]{Turok:1989ai}%
  \BibitemOpen
  \bibfield  {author} {\bibinfo {author} {\bibfnamefont {N.}~\bibnamefont
  {Turok}},\ }\href {\doibase 10.1103/PhysRevLett.63.2625} {\bibfield
  {journal} {\bibinfo  {journal} {Phys. Rev. Lett.}\ }\textbf {\bibinfo
  {volume} {63}},\ \bibinfo {pages} {2625} (\bibinfo {year}
  {1989})}\BibitemShut {NoStop}%
\bibitem [{\citenamefont {Bennett}\ and\ \citenamefont
  {Rhie}(1990)}]{Bennett:1990xy}%
  \BibitemOpen
  \bibfield  {author} {\bibinfo {author} {\bibfnamefont {D.~P.}\ \bibnamefont
  {Bennett}}\ and\ \bibinfo {author} {\bibfnamefont {S.~H.}\ \bibnamefont
  {Rhie}},\ }\href {\doibase 10.1103/PhysRevLett.65.1709} {\bibfield  {journal}
  {\bibinfo  {journal} {Phys. Rev. Lett.}\ }\textbf {\bibinfo {volume} {65}},\
  \bibinfo {pages} {1709} (\bibinfo {year} {1990})}\BibitemShut {NoStop}%
\bibitem [{\citenamefont {Zeldovich}\ \emph {et~al.}(1974)\citenamefont
  {Zeldovich}, \citenamefont {Kobzarev},\ and\ \citenamefont
  {Okun}}]{Zeldovich:1974uw}%
  \BibitemOpen
  \bibfield  {author} {\bibinfo {author} {\bibfnamefont {Y.~B.}\ \bibnamefont
  {Zeldovich}}, \bibinfo {author} {\bibfnamefont {I.~Y.}\ \bibnamefont
  {Kobzarev}}, \ and\ \bibinfo {author} {\bibfnamefont {L.~B.}\ \bibnamefont
  {Okun}},\ }\href@noop {} {\bibfield  {journal} {\bibinfo  {journal} {Zh.
  Eksp. Teor. Fiz.}\ }\textbf {\bibinfo {volume} {67}},\ \bibinfo {pages} {3}
  (\bibinfo {year} {1974})}\BibitemShut {NoStop}%
\bibitem [{\citenamefont {Abbott}\ and\ \citenamefont
  {Wise}(1989)}]{Abbott:1989jw}%
  \BibitemOpen
  \bibfield  {author} {\bibinfo {author} {\bibfnamefont {L.~F.}\ \bibnamefont
  {Abbott}}\ and\ \bibinfo {author} {\bibfnamefont {M.~B.}\ \bibnamefont
  {Wise}},\ }\href {\doibase 10.1016/0550-3213(89)90503-8} {\bibfield
  {journal} {\bibinfo  {journal} {Nucl. Phys. B}\ }\textbf {\bibinfo {volume}
  {325}},\ \bibinfo {pages} {687} (\bibinfo {year} {1989})}\BibitemShut
  {NoStop}%
\bibitem [{\citenamefont {Gelmini}\ \emph {et~al.}(1989)\citenamefont
  {Gelmini}, \citenamefont {Gleiser},\ and\ \citenamefont
  {Kolb}}]{Gelmini:1988sf}%
  \BibitemOpen
  \bibfield  {author} {\bibinfo {author} {\bibfnamefont {G.~B.}\ \bibnamefont
  {Gelmini}}, \bibinfo {author} {\bibfnamefont {M.}~\bibnamefont {Gleiser}}, \
  and\ \bibinfo {author} {\bibfnamefont {E.~W.}\ \bibnamefont {Kolb}},\ }\href
  {\doibase 10.1103/PhysRevD.39.1558} {\bibfield  {journal} {\bibinfo
  {journal} {Phys. Rev. D}\ }\textbf {\bibinfo {volume} {39}},\ \bibinfo
  {pages} {1558} (\bibinfo {year} {1989})}\BibitemShut {NoStop}%
\bibitem [{\citenamefont {Larsson}\ \emph {et~al.}(1997)\citenamefont
  {Larsson}, \citenamefont {Sarkar},\ and\ \citenamefont
  {White}}]{Larsson:1996sp}%
  \BibitemOpen
  \bibfield  {author} {\bibinfo {author} {\bibfnamefont {S.~E.}\ \bibnamefont
  {Larsson}}, \bibinfo {author} {\bibfnamefont {S.}~\bibnamefont {Sarkar}}, \
  and\ \bibinfo {author} {\bibfnamefont {P.~L.}\ \bibnamefont {White}},\ }\href
  {\doibase 10.1103/PhysRevD.55.5129} {\bibfield  {journal} {\bibinfo
  {journal} {Phys. Rev. D}\ }\textbf {\bibinfo {volume} {55}},\ \bibinfo
  {pages} {5129} (\bibinfo {year} {1997})},\ \Eprint
  {http://arxiv.org/abs/hep-ph/9608319} {arXiv:hep-ph/9608319} \BibitemShut
  {NoStop}%
\bibitem [{\citenamefont {Chang}\ \emph {et~al.}(1999)\citenamefont {Chang},
  \citenamefont {Hagmann},\ and\ \citenamefont {Sikivie}}]{Chang:1998tb}%
  \BibitemOpen
  \bibfield  {author} {\bibinfo {author} {\bibfnamefont {S.}~\bibnamefont
  {Chang}}, \bibinfo {author} {\bibfnamefont {C.}~\bibnamefont {Hagmann}}, \
  and\ \bibinfo {author} {\bibfnamefont {P.}~\bibnamefont {Sikivie}},\ }\href
  {\doibase 10.1103/PhysRevD.59.023505} {\bibfield  {journal} {\bibinfo
  {journal} {Phys. Rev. D}\ }\textbf {\bibinfo {volume} {59}},\ \bibinfo
  {pages} {023505} (\bibinfo {year} {1999})},\ \Eprint
  {http://arxiv.org/abs/hep-ph/9807374} {arXiv:hep-ph/9807374} \BibitemShut
  {NoStop}%
\bibitem [{\citenamefont {Vilenkin}\ and\ \citenamefont
  {Shellard}(2000)}]{Vilenkin:2000jqa}%
  \BibitemOpen
  \bibfield  {author} {\bibinfo {author} {\bibfnamefont {A.}~\bibnamefont
  {Vilenkin}}\ and\ \bibinfo {author} {\bibfnamefont {E.~P.~S.}\ \bibnamefont
  {Shellard}},\ }\href@noop {} {\emph {\bibinfo {title} {{Cosmic Strings and
  Other Topological Defects}}}}\ (\bibinfo  {publisher} {Cambridge University
  Press},\ \bibinfo {year} {2000})\BibitemShut {NoStop}%
\bibitem [{\citenamefont {Jana}\ \emph {et~al.}(2026)\citenamefont {Jana},
  \citenamefont {Manna},\ and\ \citenamefont {K}}]{Jana:2025vyb}%
  \BibitemOpen
  \bibfield  {author} {\bibinfo {author} {\bibfnamefont {S.}~\bibnamefont
  {Jana}}, \bibinfo {author} {\bibfnamefont {S.}~\bibnamefont {Manna}}, \ and\
  \bibinfo {author} {\bibfnamefont {V.~P.}\ \bibnamefont {K}},\ }\href
  {\doibase 10.1016/j.physletb.2026.140476} {\bibfield  {journal} {\bibinfo
  {journal} {Phys. Lett. B}\ }\textbf {\bibinfo {volume} {877}},\ \bibinfo
  {pages} {140476} (\bibinfo {year} {2026})},\ \Eprint
  {http://arxiv.org/abs/2509.10456} {arXiv:2509.10456 [hep-ph]} \BibitemShut
  {NoStop}%
\bibitem [{\citenamefont {Ai}(2026)}]{Ai:2026zpy}%
  \BibitemOpen
  \bibfield  {author} {\bibinfo {author} {\bibfnamefont {W.-Y.}\ \bibnamefont
  {Ai}},\ }\href@noop {} {\  (\bibinfo {year} {2026})},\ \Eprint
  {http://arxiv.org/abs/2606.30094} {arXiv:2606.30094 [hep-ph]} \BibitemShut
  {NoStop}%
\bibitem [{\citenamefont {Zhang}\ \emph {et~al.}(2023)\citenamefont {Zhang},
  \citenamefont {Cai}, \citenamefont {Su}, \citenamefont {Wang}, \citenamefont
  {Yu},\ and\ \citenamefont {Zhang}}]{Zhang:2023nrs}%
  \BibitemOpen
  \bibfield  {author} {\bibinfo {author} {\bibfnamefont {Z.}~\bibnamefont
  {Zhang}}, \bibinfo {author} {\bibfnamefont {C.}~\bibnamefont {Cai}}, \bibinfo
  {author} {\bibfnamefont {Y.-H.}\ \bibnamefont {Su}}, \bibinfo {author}
  {\bibfnamefont {S.}~\bibnamefont {Wang}}, \bibinfo {author} {\bibfnamefont
  {Z.-H.}\ \bibnamefont {Yu}}, \ and\ \bibinfo {author} {\bibfnamefont {H.-H.}\
  \bibnamefont {Zhang}},\ }\href {\doibase 10.1103/PhysRevD.108.095037}
  {\bibfield  {journal} {\bibinfo  {journal} {Phys. Rev. D}\ }\textbf {\bibinfo
  {volume} {108}},\ \bibinfo {pages} {095037} (\bibinfo {year} {2023})},\
  \Eprint {http://arxiv.org/abs/2307.11495} {arXiv:2307.11495 [hep-ph]}
  \BibitemShut {NoStop}%
\bibitem [{\citenamefont {Zeng}\ \emph {et~al.}(2025)\citenamefont {Zeng},
  \citenamefont {He}, \citenamefont {Yu},\ and\ \citenamefont
  {Zheng}}]{Zeng:2025zjp}%
  \BibitemOpen
  \bibfield  {author} {\bibinfo {author} {\bibfnamefont {Q.-Q.}\ \bibnamefont
  {Zeng}}, \bibinfo {author} {\bibfnamefont {X.}~\bibnamefont {He}}, \bibinfo
  {author} {\bibfnamefont {Z.-H.}\ \bibnamefont {Yu}}, \ and\ \bibinfo {author}
  {\bibfnamefont {J.}~\bibnamefont {Zheng}},\ }\href {\doibase
  10.1103/cdsj-bmvx} {\bibfield  {journal} {\bibinfo  {journal} {Phys. Rev. D}\
  }\textbf {\bibinfo {volume} {111}},\ \bibinfo {pages} {115017} (\bibinfo
  {year} {2025})},\ \Eprint {http://arxiv.org/abs/2501.10059} {arXiv:2501.10059
  [hep-ph]} \BibitemShut {NoStop}%
\bibitem [{\citenamefont {Borah}\ and\ \citenamefont
  {Saha}(2025)}]{Borah:2025bfa}%
  \BibitemOpen
  \bibfield  {author} {\bibinfo {author} {\bibfnamefont {D.}~\bibnamefont
  {Borah}}\ and\ \bibinfo {author} {\bibfnamefont {I.}~\bibnamefont {Saha}},\
  }\href@noop {} {\  (\bibinfo {year} {2025})},\ \Eprint
  {http://arxiv.org/abs/2512.22339} {arXiv:2512.22339 [hep-ph]} \BibitemShut
  {NoStop}%
\bibitem [{\citenamefont {Borah}\ \emph {et~al.}(2026)\citenamefont {Borah},
  \citenamefont {Paul},\ and\ \citenamefont {Sahu}}]{Borah:2026kfo}%
  \BibitemOpen
  \bibfield  {author} {\bibinfo {author} {\bibfnamefont {D.}~\bibnamefont
  {Borah}}, \bibinfo {author} {\bibfnamefont {P.~K.}\ \bibnamefont {Paul}}, \
  and\ \bibinfo {author} {\bibfnamefont {N.}~\bibnamefont {Sahu}},\ }\href@noop
  {} {\  (\bibinfo {year} {2026})},\ \Eprint {http://arxiv.org/abs/2602.07380}
  {arXiv:2602.07380 [hep-ph]} \BibitemShut {NoStop}%
\bibitem [{\citenamefont {Bhandari}\ \emph {et~al.}(2026)\citenamefont
  {Bhandari}, \citenamefont {Borah},\ and\ \citenamefont
  {Saha}}]{Bhandari:2026ujy}%
  \BibitemOpen
  \bibfield  {author} {\bibinfo {author} {\bibfnamefont {D.}~\bibnamefont
  {Bhandari}}, \bibinfo {author} {\bibfnamefont {D.}~\bibnamefont {Borah}}, \
  and\ \bibinfo {author} {\bibfnamefont {I.}~\bibnamefont {Saha}},\ }\href@noop
  {} {\  (\bibinfo {year} {2026})},\ \Eprint {http://arxiv.org/abs/2604.02421}
  {arXiv:2604.02421 [hep-ph]} \BibitemShut {NoStop}%
\bibitem [{\citenamefont {Zhang}(2024)}]{Zhang:2023gfu}%
  \BibitemOpen
  \bibfield  {author} {\bibinfo {author} {\bibfnamefont {Y.}~\bibnamefont
  {Zhang}},\ }\href {\doibase 10.1103/PhysRevLett.132.081003} {\bibfield
  {journal} {\bibinfo  {journal} {Phys. Rev. Lett.}\ }\textbf {\bibinfo
  {volume} {132}},\ \bibinfo {pages} {081003} (\bibinfo {year} {2024})},\
  \Eprint {http://arxiv.org/abs/2305.15495} {arXiv:2305.15495 [hep-ph]}
  \BibitemShut {NoStop}%
\bibitem [{\citenamefont {Chikashige}\ \emph {et~al.}(1981)\citenamefont
  {Chikashige}, \citenamefont {Mohapatra},\ and\ \citenamefont
  {Peccei}}]{Chikashige:1980ui}%
  \BibitemOpen
  \bibfield  {author} {\bibinfo {author} {\bibfnamefont {Y.}~\bibnamefont
  {Chikashige}}, \bibinfo {author} {\bibfnamefont {R.~N.}\ \bibnamefont
  {Mohapatra}}, \ and\ \bibinfo {author} {\bibfnamefont {R.~D.}\ \bibnamefont
  {Peccei}},\ }\href {\doibase 10.1016/0370-2693(81)90011-3} {\bibfield
  {journal} {\bibinfo  {journal} {Phys. Lett. B}\ }\textbf {\bibinfo {volume}
  {98}},\ \bibinfo {pages} {265} (\bibinfo {year} {1981})}\BibitemShut
  {NoStop}%
\bibitem [{\citenamefont {Minkowski}(1977)}]{Minkowski:1977sc}%
  \BibitemOpen
  \bibfield  {author} {\bibinfo {author} {\bibfnamefont {P.}~\bibnamefont
  {Minkowski}},\ }\href {\doibase 10.1016/0370-2693(77)90435-X} {\bibfield
  {journal} {\bibinfo  {journal} {Phys. Lett. B}\ }\textbf {\bibinfo {volume}
  {67}},\ \bibinfo {pages} {421} (\bibinfo {year} {1977})}\BibitemShut
  {NoStop}%
\bibitem [{\citenamefont {Yanagida}(1980)}]{Yanagida:1980xy}%
  \BibitemOpen
  \bibfield  {author} {\bibinfo {author} {\bibfnamefont {T.}~\bibnamefont
  {Yanagida}},\ }\href {\doibase 10.1143/PTP.64.1103} {\bibfield  {journal}
  {\bibinfo  {journal} {Prog. Theor. Phys.}\ }\textbf {\bibinfo {volume}
  {64}},\ \bibinfo {pages} {1103} (\bibinfo {year} {1980})}\BibitemShut
  {NoStop}%
\bibitem [{\citenamefont {Gell-Mann}\ \emph
  {et~al.}(1979{\natexlab{a}})\citenamefont {Gell-Mann}, \citenamefont
  {Ramond},\ and\ \citenamefont {Slansky}}]{Gell-Mann:1979ijt}%
  \BibitemOpen
  \bibfield  {author} {\bibinfo {author} {\bibfnamefont {M.}~\bibnamefont
  {Gell-Mann}}, \bibinfo {author} {\bibfnamefont {P.}~\bibnamefont {Ramond}}, \
  and\ \bibinfo {author} {\bibfnamefont {R.}~\bibnamefont {Slansky}},\ }in\
  \href@noop {} {\emph {\bibinfo {booktitle} {{International Symposium on
  Fundamentals of Quantum Theory and Quantum Field Theory}}}}\ (\bibinfo {year}
  {1979})\ \Eprint {http://arxiv.org/abs/hep-ph/9809459} {arXiv:hep-ph/9809459}
  \BibitemShut {NoStop}%
\bibitem [{\citenamefont {Gell-Mann}\ \emph
  {et~al.}(1979{\natexlab{b}})\citenamefont {Gell-Mann}, \citenamefont
  {Ramond},\ and\ \citenamefont {Slansky}}]{Gell-Mann:1979vob}%
  \BibitemOpen
  \bibfield  {author} {\bibinfo {author} {\bibfnamefont {M.}~\bibnamefont
  {Gell-Mann}}, \bibinfo {author} {\bibfnamefont {P.}~\bibnamefont {Ramond}}, \
  and\ \bibinfo {author} {\bibfnamefont {R.}~\bibnamefont {Slansky}},\
  }\href@noop {} {\bibfield  {journal} {\bibinfo  {journal} {Conf. Proc. C}\
  }\textbf {\bibinfo {volume} {790927}},\ \bibinfo {pages} {315} (\bibinfo
  {year} {1979}{\natexlab{b}})},\ \Eprint {http://arxiv.org/abs/1306.4669}
  {arXiv:1306.4669 [hep-th]} \BibitemShut {NoStop}%
\bibitem [{\citenamefont {Glashow}(1980)}]{Glashow:1979nm}%
  \BibitemOpen
  \bibfield  {author} {\bibinfo {author} {\bibfnamefont {S.~L.}\ \bibnamefont
  {Glashow}},\ }\href {\doibase 10.1007/978-1-4684-7197-7_15} {\bibfield
  {journal} {\bibinfo  {journal} {NATO Sci. Ser. B}\ }\textbf {\bibinfo
  {volume} {61}},\ \bibinfo {pages} {687} (\bibinfo {year} {1980})}\BibitemShut
  {NoStop}%
\bibitem [{\citenamefont {Mohapatra}\ and\ \citenamefont
  {Senjanovic}(1980)}]{Mohapatra:1979ia}%
  \BibitemOpen
  \bibfield  {author} {\bibinfo {author} {\bibfnamefont {R.~N.}\ \bibnamefont
  {Mohapatra}}\ and\ \bibinfo {author} {\bibfnamefont {G.}~\bibnamefont
  {Senjanovic}},\ }\href {\doibase 10.1103/PhysRevLett.44.912} {\bibfield
  {journal} {\bibinfo  {journal} {Phys. Rev. Lett.}\ }\textbf {\bibinfo
  {volume} {44}},\ \bibinfo {pages} {912} (\bibinfo {year} {1980})}\BibitemShut
  {NoStop}%
\bibitem [{\citenamefont {Schechter}\ and\ \citenamefont
  {Valle}(1980)}]{Schechter:1980gr}%
  \BibitemOpen
  \bibfield  {author} {\bibinfo {author} {\bibfnamefont {J.}~\bibnamefont
  {Schechter}}\ and\ \bibinfo {author} {\bibfnamefont {J.~W.~F.}\ \bibnamefont
  {Valle}},\ }\href {\doibase 10.1103/PhysRevD.22.2227} {\bibfield  {journal}
  {\bibinfo  {journal} {Phys. Rev. D}\ }\textbf {\bibinfo {volume} {22}},\
  \bibinfo {pages} {2227} (\bibinfo {year} {1980})}\BibitemShut {NoStop}%
\bibitem [{\citenamefont {Georgi}\ \emph {et~al.}(1981)\citenamefont {Georgi},
  \citenamefont {Glashow},\ and\ \citenamefont {Nussinov}}]{Georgi:1981pg}%
  \BibitemOpen
  \bibfield  {author} {\bibinfo {author} {\bibfnamefont {H.~M.}\ \bibnamefont
  {Georgi}}, \bibinfo {author} {\bibfnamefont {S.~L.}\ \bibnamefont {Glashow}},
  \ and\ \bibinfo {author} {\bibfnamefont {S.}~\bibnamefont {Nussinov}},\
  }\href {\doibase 10.1016/0550-3213(81)90336-9} {\bibfield  {journal}
  {\bibinfo  {journal} {Nucl. Phys. B}\ }\textbf {\bibinfo {volume} {193}},\
  \bibinfo {pages} {297} (\bibinfo {year} {1981})}\BibitemShut {NoStop}%
\bibitem [{\citenamefont {Preskill}(1992)}]{Preskill:1992tc}%
  \BibitemOpen
  \bibfield  {author} {\bibinfo {author} {\bibfnamefont {J.}~\bibnamefont
  {Preskill}},\ }in\ \href@noop {} {\emph {\bibinfo {booktitle} {{International
  Symposium on Black holes, Membranes, Wormholes and Superstrings}}}}\
  (\bibinfo {year} {1992})\ \Eprint {http://arxiv.org/abs/hep-th/9209058}
  {arXiv:hep-th/9209058} \BibitemShut {NoStop}%
\bibitem [{\citenamefont {Greljo}\ \emph {et~al.}(2025)\citenamefont {Greljo},
  \citenamefont {Ponce~D{\'\i}az},\ and\ \citenamefont
  {Thomsen}}]{Greljo:2025suh}%
  \BibitemOpen
  \bibfield  {author} {\bibinfo {author} {\bibfnamefont {A.}~\bibnamefont
  {Greljo}}, \bibinfo {author} {\bibfnamefont {X.}~\bibnamefont
  {Ponce~D{\'\i}az}}, \ and\ \bibinfo {author} {\bibfnamefont {A.~E.}\
  \bibnamefont {Thomsen}},\ }\href {\doibase 10.1088/1475-7516/2025/11/043}
  {\bibfield  {journal} {\bibinfo  {journal} {JCAP}\ }\textbf {\bibinfo
  {volume} {11}},\ \bibinfo {pages} {043} (\bibinfo {year} {2025})},\ \Eprint
  {http://arxiv.org/abs/2505.18259} {arXiv:2505.18259 [hep-ph]} \BibitemShut
  {NoStop}%
\bibitem [{\citenamefont {Ghosh}\ \emph {et~al.}(2026)\citenamefont {Ghosh},
  \citenamefont {Loho},\ and\ \citenamefont {Manna}}]{Ghosh:2025cxp}%
  \BibitemOpen
  \bibfield  {author} {\bibinfo {author} {\bibfnamefont {T.}~\bibnamefont
  {Ghosh}}, \bibinfo {author} {\bibfnamefont {K.}~\bibnamefont {Loho}}, \ and\
  \bibinfo {author} {\bibfnamefont {S.}~\bibnamefont {Manna}},\ }\href
  {\doibase 10.1103/gd2q-qdhg} {\bibfield  {journal} {\bibinfo  {journal}
  {Phys. Rev. D}\ }\textbf {\bibinfo {volume} {113}},\ \bibinfo {pages}
  {043036} (\bibinfo {year} {2026})},\ \Eprint
  {http://arxiv.org/abs/2507.04342} {arXiv:2507.04342 [hep-ph]} \BibitemShut
  {NoStop}%
\bibitem [{\citenamefont {Preskill}\ and\ \citenamefont
  {Vilenkin}(1993)}]{Preskill:1992ck}%
  \BibitemOpen
  \bibfield  {author} {\bibinfo {author} {\bibfnamefont {J.}~\bibnamefont
  {Preskill}}\ and\ \bibinfo {author} {\bibfnamefont {A.}~\bibnamefont
  {Vilenkin}},\ }\href {\doibase 10.1103/PhysRevD.47.2324} {\bibfield
  {journal} {\bibinfo  {journal} {Phys. Rev. D}\ }\textbf {\bibinfo {volume}
  {47}},\ \bibinfo {pages} {2324} (\bibinfo {year} {1993})},\ \Eprint
  {http://arxiv.org/abs/hep-ph/9209210} {arXiv:hep-ph/9209210} \BibitemShut
  {NoStop}%
\bibitem [{\citenamefont {Wu}\ \emph {et~al.}(2022{\natexlab{a}})\citenamefont
  {Wu}, \citenamefont {Xie},\ and\ \citenamefont {Zhou}}]{Wu:2022tpe}%
  \BibitemOpen
  \bibfield  {author} {\bibinfo {author} {\bibfnamefont {Y.}~\bibnamefont
  {Wu}}, \bibinfo {author} {\bibfnamefont {K.-P.}\ \bibnamefont {Xie}}, \ and\
  \bibinfo {author} {\bibfnamefont {Y.-L.}\ \bibnamefont {Zhou}},\ }\href
  {\doibase 10.1103/PhysRevD.106.075019} {\bibfield  {journal} {\bibinfo
  {journal} {Phys. Rev. D}\ }\textbf {\bibinfo {volume} {106}},\ \bibinfo
  {pages} {075019} (\bibinfo {year} {2022}{\natexlab{a}})},\ \Eprint
  {http://arxiv.org/abs/2205.11529} {arXiv:2205.11529 [hep-ph]} \BibitemShut
  {NoStop}%
\bibitem [{\citenamefont {Blasi}\ \emph
  {et~al.}(2023{\natexlab{a}})\citenamefont {Blasi}, \citenamefont {Mariotti},
  \citenamefont {Rase},\ and\ \citenamefont {Sevrin}}]{Blasi:2023sej}%
  \BibitemOpen
  \bibfield  {author} {\bibinfo {author} {\bibfnamefont {S.}~\bibnamefont
  {Blasi}}, \bibinfo {author} {\bibfnamefont {A.}~\bibnamefont {Mariotti}},
  \bibinfo {author} {\bibfnamefont {A.}~\bibnamefont {Rase}}, \ and\ \bibinfo
  {author} {\bibfnamefont {A.}~\bibnamefont {Sevrin}},\ }\href {\doibase
  10.1007/JHEP11(2023)169} {\bibfield  {journal} {\bibinfo  {journal} {JHEP}\
  }\textbf {\bibinfo {volume} {11}},\ \bibinfo {pages} {169} (\bibinfo {year}
  {2023}{\natexlab{a}})},\ \Eprint {http://arxiv.org/abs/2306.17830}
  {arXiv:2306.17830 [hep-ph]} \BibitemShut {NoStop}%
\bibitem [{\citenamefont {Berbig}(2025)}]{Berbig:2025nrt}%
  \BibitemOpen
  \bibfield  {author} {\bibinfo {author} {\bibfnamefont {M.}~\bibnamefont
  {Berbig}},\ }\href@noop {} {\  (\bibinfo {year} {2025})},\ \Eprint
  {http://arxiv.org/abs/2506.02910} {arXiv:2506.02910 [hep-ph]} \BibitemShut
  {NoStop}%
\bibitem [{\citenamefont {Giddings}\ and\ \citenamefont
  {Strominger}(1988)}]{Giddings:1988cx}%
  \BibitemOpen
  \bibfield  {author} {\bibinfo {author} {\bibfnamefont {S.~B.}\ \bibnamefont
  {Giddings}}\ and\ \bibinfo {author} {\bibfnamefont {A.}~\bibnamefont
  {Strominger}},\ }\href {\doibase 10.1016/0550-3213(88)90109-5} {\bibfield
  {journal} {\bibinfo  {journal} {Nucl. Phys. B}\ }\textbf {\bibinfo {volume}
  {307}},\ \bibinfo {pages} {854} (\bibinfo {year} {1988})}\BibitemShut
  {NoStop}%
\bibitem [{\citenamefont {Heidenreich}\ \emph {et~al.}(2021)\citenamefont
  {Heidenreich}, \citenamefont {McNamara}, \citenamefont {Montero},
  \citenamefont {Reece}, \citenamefont {Rudelius},\ and\ \citenamefont
  {Valenzuela}}]{Heidenreich:2020pkc}%
  \BibitemOpen
  \bibfield  {author} {\bibinfo {author} {\bibfnamefont {B.}~\bibnamefont
  {Heidenreich}}, \bibinfo {author} {\bibfnamefont {J.}~\bibnamefont
  {McNamara}}, \bibinfo {author} {\bibfnamefont {M.}~\bibnamefont {Montero}},
  \bibinfo {author} {\bibfnamefont {M.}~\bibnamefont {Reece}}, \bibinfo
  {author} {\bibfnamefont {T.}~\bibnamefont {Rudelius}}, \ and\ \bibinfo
  {author} {\bibfnamefont {I.}~\bibnamefont {Valenzuela}},\ }\href {\doibase
  10.1007/JHEP11(2021)053} {\bibfield  {journal} {\bibinfo  {journal} {JHEP}\
  }\textbf {\bibinfo {volume} {11}},\ \bibinfo {pages} {053} (\bibinfo {year}
  {2021})},\ \Eprint {http://arxiv.org/abs/2012.00009} {arXiv:2012.00009
  [hep-th]} \BibitemShut {NoStop}%
\bibitem [{\citenamefont {Reece}(2024)}]{Reece:2023czb}%
  \BibitemOpen
  \bibfield  {author} {\bibinfo {author} {\bibfnamefont {M.}~\bibnamefont
  {Reece}},\ }\href {\doibase 10.22323/1.439.0008} {\bibfield  {journal}
  {\bibinfo  {journal} {PoS}\ }\textbf {\bibinfo {volume} {TASI2022}},\
  \bibinfo {pages} {008} (\bibinfo {year} {2024})},\ \Eprint
  {http://arxiv.org/abs/2304.08512} {arXiv:2304.08512 [hep-ph]} \BibitemShut
  {NoStop}%
\bibitem [{\citenamefont {Coleman}\ and\ \citenamefont
  {Weinberg}(1973)}]{Coleman:1973jx}%
  \BibitemOpen
  \bibfield  {author} {\bibinfo {author} {\bibfnamefont {S.~R.}\ \bibnamefont
  {Coleman}}\ and\ \bibinfo {author} {\bibfnamefont {E.~J.}\ \bibnamefont
  {Weinberg}},\ }\href {\doibase 10.1103/PhysRevD.7.1888} {\bibfield  {journal}
  {\bibinfo  {journal} {Phys. Rev. D}\ }\textbf {\bibinfo {volume} {7}},\
  \bibinfo {pages} {1888} (\bibinfo {year} {1973})}\BibitemShut {NoStop}%
\bibitem [{\citenamefont {Gelmini}\ \emph {et~al.}(2021)\citenamefont
  {Gelmini}, \citenamefont {Pascoli}, \citenamefont {Vitagliano},\ and\
  \citenamefont {Zhou}}]{Gelmini:2020bqg}%
  \BibitemOpen
  \bibfield  {author} {\bibinfo {author} {\bibfnamefont {G.~B.}\ \bibnamefont
  {Gelmini}}, \bibinfo {author} {\bibfnamefont {S.}~\bibnamefont {Pascoli}},
  \bibinfo {author} {\bibfnamefont {E.}~\bibnamefont {Vitagliano}}, \ and\
  \bibinfo {author} {\bibfnamefont {Y.-L.}\ \bibnamefont {Zhou}},\ }\href
  {\doibase 10.1088/1475-7516/2021/02/032} {\bibfield  {journal} {\bibinfo
  {journal} {JCAP}\ }\textbf {\bibinfo {volume} {02}},\ \bibinfo {pages} {032}
  (\bibinfo {year} {2021})},\ \Eprint {http://arxiv.org/abs/2009.01903}
  {arXiv:2009.01903 [hep-ph]} \BibitemShut {NoStop}%
\bibitem [{\citenamefont {Blasi}\ \emph
  {et~al.}(2023{\natexlab{b}})\citenamefont {Blasi}, \citenamefont {Mariotti},
  \citenamefont {Rase}, \citenamefont {Sevrin},\ and\ \citenamefont
  {Turbang}}]{Blasi:2022ayo}%
  \BibitemOpen
  \bibfield  {author} {\bibinfo {author} {\bibfnamefont {S.}~\bibnamefont
  {Blasi}}, \bibinfo {author} {\bibfnamefont {A.}~\bibnamefont {Mariotti}},
  \bibinfo {author} {\bibfnamefont {A.}~\bibnamefont {Rase}}, \bibinfo {author}
  {\bibfnamefont {A.}~\bibnamefont {Sevrin}}, \ and\ \bibinfo {author}
  {\bibfnamefont {K.}~\bibnamefont {Turbang}},\ }\href {\doibase
  10.1088/1475-7516/2023/04/008} {\bibfield  {journal} {\bibinfo  {journal}
  {JCAP}\ }\textbf {\bibinfo {volume} {04}},\ \bibinfo {pages} {008} (\bibinfo
  {year} {2023}{\natexlab{b}})},\ \Eprint {http://arxiv.org/abs/2210.14246}
  {arXiv:2210.14246 [hep-ph]} \BibitemShut {NoStop}%
\bibitem [{\citenamefont {Wu}\ \emph {et~al.}(2022{\natexlab{b}})\citenamefont
  {Wu}, \citenamefont {Xie},\ and\ \citenamefont {Zhou}}]{Wu:2022stu}%
  \BibitemOpen
  \bibfield  {author} {\bibinfo {author} {\bibfnamefont {Y.}~\bibnamefont
  {Wu}}, \bibinfo {author} {\bibfnamefont {K.-P.}\ \bibnamefont {Xie}}, \ and\
  \bibinfo {author} {\bibfnamefont {Y.-L.}\ \bibnamefont {Zhou}},\ }\href
  {\doibase 10.1103/PhysRevD.105.095013} {\bibfield  {journal} {\bibinfo
  {journal} {Phys. Rev. D}\ }\textbf {\bibinfo {volume} {105}},\ \bibinfo
  {pages} {095013} (\bibinfo {year} {2022}{\natexlab{b}})},\ \Eprint
  {http://arxiv.org/abs/2204.04374} {arXiv:2204.04374 [hep-ph]} \BibitemShut
  {NoStop}%
\bibitem [{\citenamefont {Li}\ \emph {et~al.}(2025)\citenamefont {Li},
  \citenamefont {Liu},\ and\ \citenamefont {Guo}}]{Li:2025gld}%
  \BibitemOpen
  \bibfield  {author} {\bibinfo {author} {\bibfnamefont {Y.-J.}\ \bibnamefont
  {Li}}, \bibinfo {author} {\bibfnamefont {J.}~\bibnamefont {Liu}}, \ and\
  \bibinfo {author} {\bibfnamefont {Z.-K.}\ \bibnamefont {Guo}},\ }\href
  {\doibase 10.1103/rnpp-7wh2} {\bibfield  {journal} {\bibinfo  {journal}
  {Phys. Rev. D}\ }\textbf {\bibinfo {volume} {112}},\ \bibinfo {pages}
  {103510} (\bibinfo {year} {2025})},\ \Eprint
  {http://arxiv.org/abs/2502.13644} {arXiv:2502.13644 [astro-ph.CO]}
  \BibitemShut {NoStop}%
\bibitem [{\citenamefont {Hiramatsu}\ \emph {et~al.}(2013)\citenamefont
  {Hiramatsu}, \citenamefont {Kawasaki}, \citenamefont {Saikawa},\ and\
  \citenamefont {Sekiguchi}}]{Hiramatsu:2012sc}%
  \BibitemOpen
  \bibfield  {author} {\bibinfo {author} {\bibfnamefont {T.}~\bibnamefont
  {Hiramatsu}}, \bibinfo {author} {\bibfnamefont {M.}~\bibnamefont {Kawasaki}},
  \bibinfo {author} {\bibfnamefont {K.}~\bibnamefont {Saikawa}}, \ and\
  \bibinfo {author} {\bibfnamefont {T.}~\bibnamefont {Sekiguchi}},\ }\href
  {\doibase 10.1088/1475-7516/2013/01/001} {\bibfield  {journal} {\bibinfo
  {journal} {JCAP}\ }\textbf {\bibinfo {volume} {01}},\ \bibinfo {pages} {001}
  (\bibinfo {year} {2013})},\ \Eprint {http://arxiv.org/abs/1207.3166}
  {arXiv:1207.3166 [hep-ph]} \BibitemShut {NoStop}%
\end{thebibliography}%

\end{document}